\documentclass[useAMS,usenatbib,letters]{mn2e}
\usepackage{graphicx}

\title[$B$--Mode contamination from 3-years WMAP data]
      {$B$--Mode contamination by synchrotron emission from 
        3-years WMAP data}
\author[E. Carretti, et al.]
{E.~Carretti$^{1}$\thanks{E-mail: carretti@ira.inaf.it}, 
G.~Bernardi$^{2}$, and
S.~Cortiglioni$^{3}$\\
$^{1}$INAF -- Istituto di Radioastronomia, Via Gobetti 101, I-40129 Bologna, Italy\\
$^{2}$Kapteyn Astronomical Institute, University of Groningen,
      PO BOX 800, 9700, AV Groningen, The Netherlands\\
$^{3}$INAF -- IASF Bologna, Via Gobetti 101, Bologna, I-40129, Italy}
\begin{document}

\date{Accepted xx xx xx. Received yy yy yy; in original form zz zz zz}

\pagerange{\pageref{firstpage}--\pageref{lastpage}} \pubyear{2006}

\maketitle

\label{firstpage}

\begin{abstract}
We study the contamination of the $B$--mode of the Cosmic Microwave Background 
Polarization (CMBP) by Galactic synchrotron in the lowest emission regions
of the sky. 
The 22.8-GHz polarization map of the 3-years WMAP data release
is used to identify and analyse such regions. 
Two areas are selected with signal-to-noise ratio $S/N<2$ and $S/N<3$, 
covering $\sim 16$\% and $\sim 26$\% fraction of the sky, respectively.
The polarization power spectra of these two areas are 
dominated by the sky signal on 
large angular scales (multipoles $\ell < 15$), 
while the noise prevails on degree scales. 
Angular extrapolations show that the synchrotron emission competes 
with the CMBP $B$--mode signal for tensor-to-scalar perturbation power ratio 
$T/S = 10^{-3}$--$10^{-2}$ at 70-GHz in the 16\% lowest 
emission sky ($S/N<2$ area). These values worsen by a factor $\sim 5$ 
in the $S/N<3$ region.
The novelty is that our estimates regard the whole lowest emission regions 
and outline a contamination better than that of 
the whole high Galactic latitude sky found by the WMAP team 
($T/S>0.3$). Such regions allow $T/S \sim 10^{-3}$ to be measured directly 
which approximately corresponds to the limit imposed by using a sky coverage 
of 15\%. This opens interesting perspectives to investigate the 
inflationary model space in lowest emission regions.
\end{abstract}


\section{Introduction}\label{introSec}

The Cosmic Microwave Background Polarization (CMBP) 
allows the study of the first stages of the Universe and is 
one of the hot topics of cosmology.
One of its components, the $B$--mode, is sensitive to the primordial 
Gravitational Wave (GW) background left by inflation giving a way
to investigate the physics of the very early Universe (e.g. \citealt{ka98}).
In fact, its power spectrum on degree scales 
has a linear dependence on the tensor-to-scalar
perturbation power ratio $T/S$, which measures the amount of 
primordial GW (e.g. \citealt{boyle06,kinney06}). 
In combination with parameters measured by the CMB Temperature spectrum 
(the scalar perturbation spectral index $n$ and its running $dn / d\ln k$),
it allows us to distinguish among the numerous models of inflation.

The $B$-mode signal is faint and only upper limits 
of $T/S$ have been set so far. \citet{seljak06} and \citet{martin06}, 
using different analyses, find consistent results of $T/S < 0.22$ and 
$T/S < 0.21$ (95\% C.L.), respectively.
On the other hand, the plethora of inflation models 
does not help constrain $T/S$. This can vary
by orders of magnitude and can be even smaller than $10^{-4}$, although
\citet{boyle06} show that only models with high fine tuning degree can feature
$T/S < 10^{-3}$. 
Correspondingly, the peak $B$-mode signal can vary
from $\sim200$~nK of the present upper limits down to values even smaller than 
$3$~nK for $T/S < 10^{-4}$.

\begin{figure}
\centering
  \includegraphics[angle=90, width=1.0\hsize]{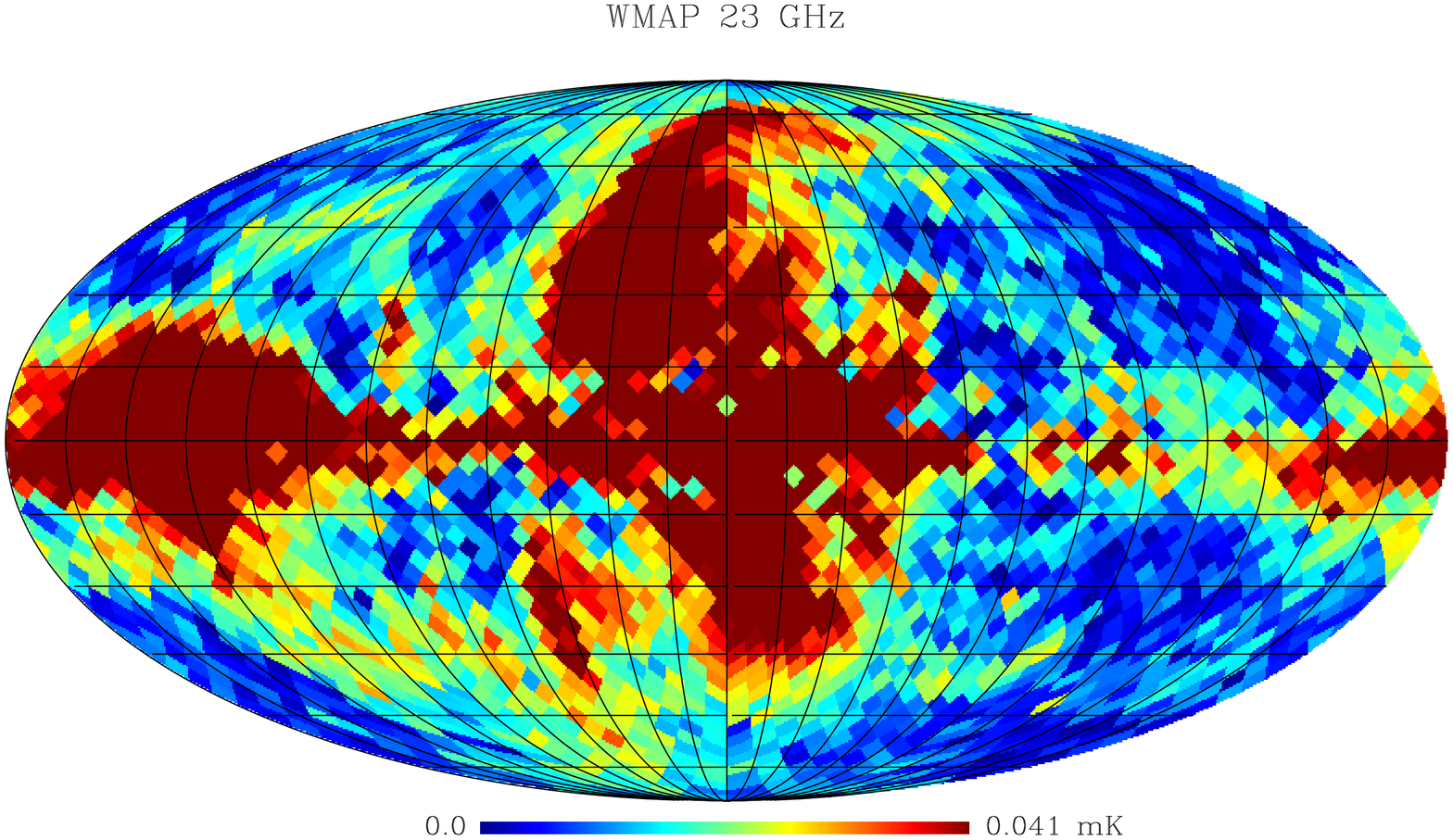}
  \includegraphics[angle=90, width=1.0\hsize]{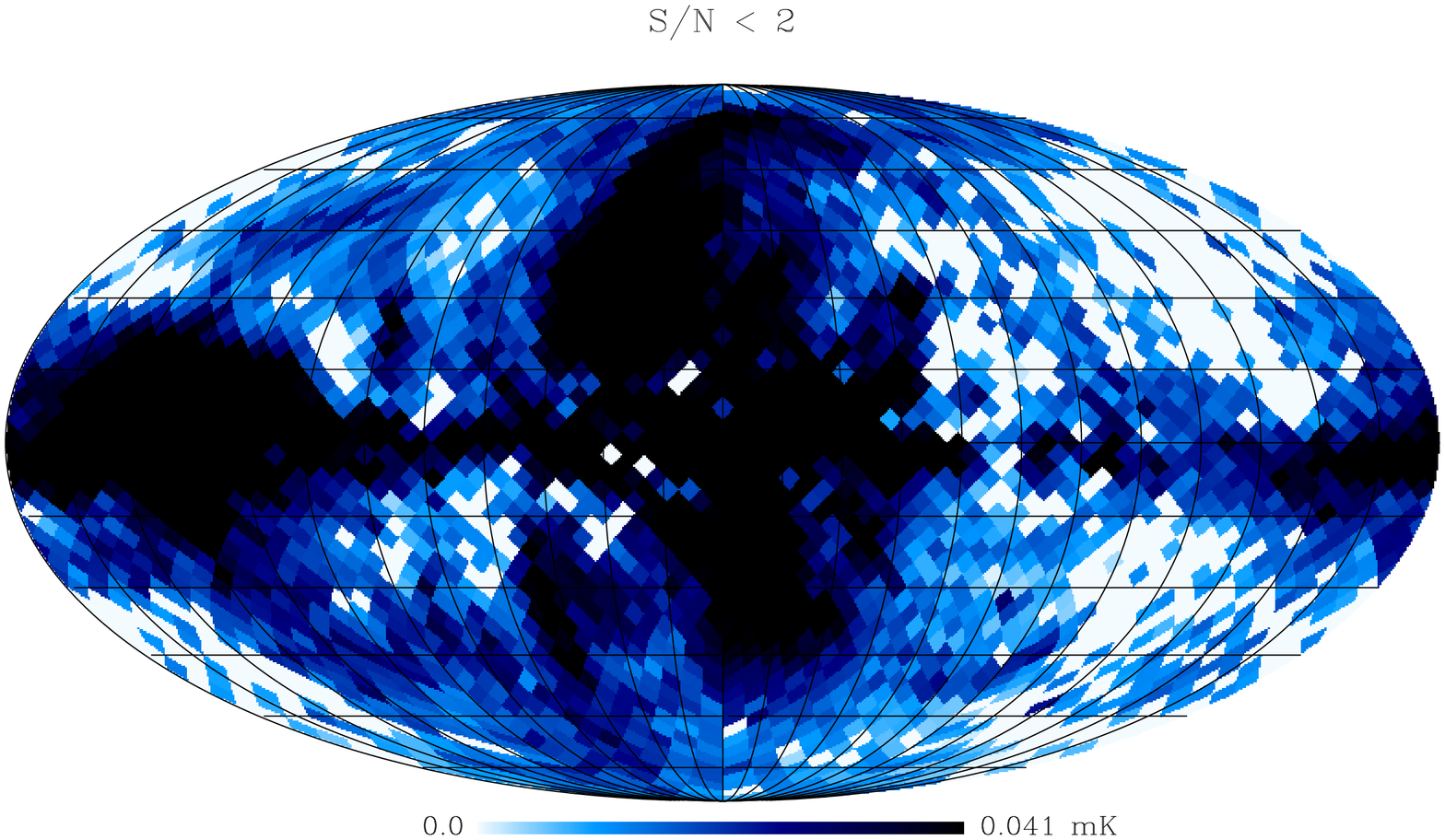}
  \includegraphics[angle=90, width=1.0\hsize]{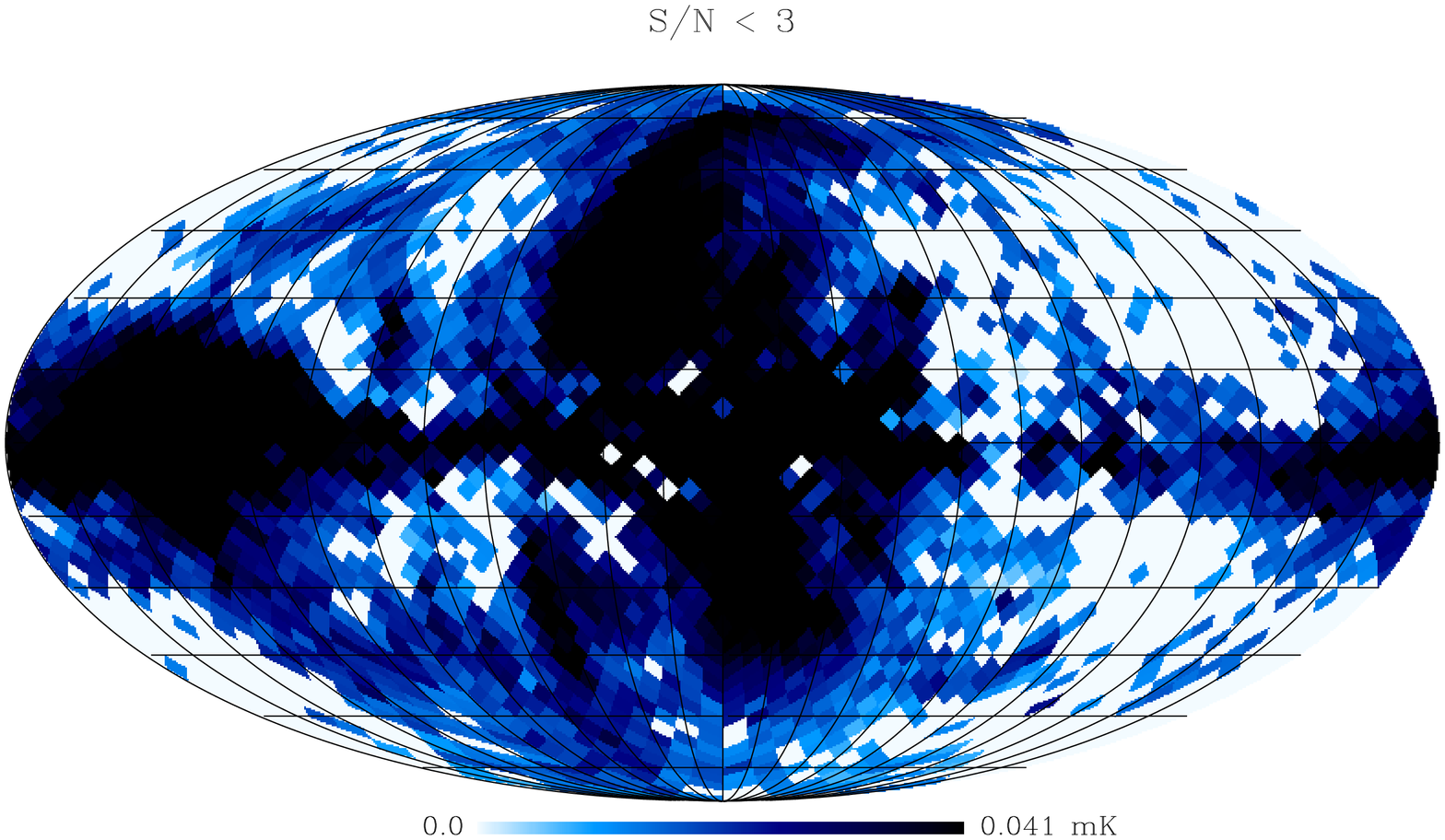}
\caption{{\it Top.} Polarized intensity map at 22.8-GHz of the WMAP experiment 
         \citep{page06}. Data are averaged on $\sim 4^\circ$ scale
         using the HEALPix \citep{gorski05} pixelation with parameter 
         $N\!side = 16$ (resolution r4). The map is in Galactic coordinates with
         the Galactic centre in the middle. The coordinate grid is $15^\circ$ 
         spaced both in longitude and latitude.
         Lowest emission regions are the blue spots at mid-high Galactic latitudes
         out of the large SNR loops. {\it Mid and bottom.} 
         Lowest emission regions in the 22.8-GHz WMAP polarization map: the 
         $4^\circ$-pixels of the two regions with $S/N < 2$ (mid) and $S/N < 3$
         (bottom) are blanked (white). 
\label{WMAPFig}
}
\end{figure}

The weakness of the CMB $B$-mode makes
it easily contaminated by foreground emissions from both the Galaxy
and extragalactic sources. 
The study of astrophysical foregrounds is thus crucial
to set the capability of CMBP experiments to investigate the early Universe and help
set which part of the inflationary model space is accessible.

At frequencies lower than 60--70~GHz, the most relevant
contaminant is expected to be the Galactic synchrotron emission.
Recent results based on 3-years Wilkinson Microwave Anisotropy Probe (WMAP) 
data at 22.8-GHz show that even the high Galactic latitudes
are normally strongly contaminated \citep{page06}. 
These authors use about 75~percent of the sky (all high latitudes but 
strong large local structures 
like the Northern Galactic Spur) and
find that at 60--70~GHz
the synchrotron emission competes with the cosmic $B$-mode signal 
even for models with $T/S = 0.3$--0.5, which are already 
disfavoured by the present upper limits.
Similar results are obtained by \citet{laporta06} 
through the analysis of the 1.4~GHz sky at latitude $|b| > 20^\circ$
using the DRAO survey data \citep{wolleben06}.

\citet{laporta06} also analyse areas\footnote{
  Three areas defined as follows: 
  A) $180^\circ<l<276^\circ$, $b>45^\circ$;
  B) $193^\circ<l<228^\circ$, $b<-45^\circ$;
  C) $65^\circ<l<180^\circ$, $b>45^\circ$;
  $l$ and $b$ are Galactic longitude and latitude.
} 
with smaller extension 
(for a total of about 10\% of the sky) 
and lower emission, finding that the Galactic signal 
at 70-GHz is comparable to the 
CMB $B$-mode in case of $T/S \sim 0.1$.
Even though lower, this signal would largely 
contaminate the CMB $B$--mode for most of the $T/S$ values presently allowed. 
However, these areas are not in the
lowest emission regions visible in the 22.8-GHz WMAP polarization map.

Information about the lowest emission regions are available, instead,
from three independent small areas observed at 1.4 and 2.3-GHz
\citep{carretti05b,carretti06,bernardi06}.
The synchrotron emission here actually looks weaker,
competing with the CMB $B$-mode only for cosmological models with 
$T/S = 10^{-3}$--$10^{-2}$ at 70-GHz \citep{carretti06}. 

Such values would give more chances to 
investigate a large part of the inflationary
model space, opening interesting perspective for the study of the 
inflation physics. 
However, the areas observed so far cover just few square degrees 
($\sim 20$-deg$^2$ in total) and represent few sparse samples of the lowest
emission regions.
A real knowledge of the typical Galactic synchrotron contribution 
of such regions is still an open question, and 
requires wider areas to be analysed.

This paper reports the first study carried out 
to identify and analyse the areas with the lowest synchrotron emission 
at microwave wavelengths, which allow negligible Faraday rotation effects (still
appreciable at $\sim 1$-GHz; e.g. see \citealt{carretti05a}).
We use the 22.8-GHz polarization map of the 3-years 
WMAP data release
to select them (Section~\ref{WMAPSec}), and compute 
their polarized angular power spectra (Section~\ref{specSec}). 
Finally, we estimate the contamination of the CMBP by Galactic synchrotron
in the CMBP frequency window (70--90~GHz) and discuss implications
for the detection of the cosmic $B$--mode in lowest emission regions 
(Section~\ref{CMBSec}).

\section{Lowest emission regions selection}\label{WMAPSec}

The 3-years WMAP data release\footnote{http://lambda.gsfc.nasa.gov/}
has made available the first polarization 
all-sky map at microwave frequencies, namely 22.8-GHz \citep{page06}.
The sensitivity does not allow to detect the signal out of the Galactic plane 
at the nominal resolution (FWHM~$\sim 1^\circ$). The high Galactic 
latitudes have low signal-to-noise ratios ($S/N$), except in large local structures
like the big Supernova Remnant (SNR) loops. 
The situation improves once data are averaged on pixels 
of $\sim 4^\circ$ (top panel of Figure~\ref{WMAPFig}), as presented by the WMAP team \citep{page06}. 
At this resolution the signal appears even at high Galactic latitudes and
allows us to identify the lowest emission regions in the sky (blue areas in
top panel of Figure~\ref{WMAPFig}), which, however, still have signal competing with the noise. 

We quantitatively select the lowest emission regions using the $4^\circ$ 
polarized intensity map. 
After accounting for the noise bias \citep{uyaniker02},
pixels with $S/N$ smaller than a given threshold have been selected.
Namely, we have considered two values: $S/N < 2$ and $S/N < 3$. 
The two resulting regions are shown in Figure~\ref{WMAPFig}.

These regions are not contiguous and located at mid-high Galactic latitudes, 
mainly in the third Galactic quadrant, 
with significant extensions in the second one. It is worth noting
that the Galactic caps are not included, having too high emission. 

The sky fractions covered are 16.2\% ($S/N < 2$) and 25.7\% ($S/N < 3$), respectively. 
Although smaller than half a sky, 
these two regions have significant extensions 
and could still effectively be used for CMBP $B$-mode studies.

\section{Power spectrum analysis}\label{specSec}

We compute the $E$-- and $B$--mode power spectra 
of the polarized emission in the two selected areas. 
These are the quantities predicted by cosmological models
and allow a direct comparison with the CMBP signal.

To account for the irregular sky coverage we use the method
based on two-point 
correlation functions of the Stokes parameters $Q$ and $U$  
described by \citet{sbarra03}. 
Correlation functions are estimated on the $Q$ and $U$ maps 
of the selected regions as   
\begin{equation}     
 \tilde{C}^X(\theta) = X_i X_j \hspace{1cm} X =    
 Q,U
\end{equation}     
where $X_i$ is the pixel $i$ content of map $X$, $i$ and $j$    
identify pixel pairs at distance $\theta$.
Data are binned with pixel-size resolution.
Power spectra $C^{E,B}_\ell$ are obtained by integration 
\begin{equation}      
  C^E_\ell+C^{E,n}_\ell = W_\ell 
              \int^\pi_0 [\tilde{C}^Q(\theta)F_{1,\ell 2}(\theta) +     
                          \tilde{C}^U(\theta)F_{2,\ell 2}(\theta)]   
                          \sin(\theta)d \theta
\end{equation}      
\begin{equation}      
 C^B_\ell+C^{B,n}_\ell  = W_\ell 
              \int^\pi_0 [\tilde{C}^U(\theta)F_{1,\ell 2}(\theta) +     
                           \tilde{C}^Q(\theta)F_{2,\ell 2}(\theta)]   
                           \sin(\theta)d \theta 
\end{equation}      
where the functions $F_{1,\ell m}$ and $F_{2,\ell m}$ are described by 
\citet{zaldarriaga98}, 
$C_\ell^{E,{\rm n}}$ and $C_\ell^{B,{\rm n}}$ are the noise spectra, and 
$W_\ell$ is the pixel window function accounting for pixel smearing effects.
The WMAP map smoothed on $1^\circ$ pixels (HEALPix parameter $Nside=64$) 
has been used to compute correlation functions and spectra, 
instead of the $4^\circ$ one used for the selection. 
This allows us to reduce the effects of the pixel window function and 
provides reliable computation up to $\ell\sim 100$ of the 
$B$--mode peak.

\begin{figure}
\centering
  \includegraphics[angle=0, width=1.0\hsize]{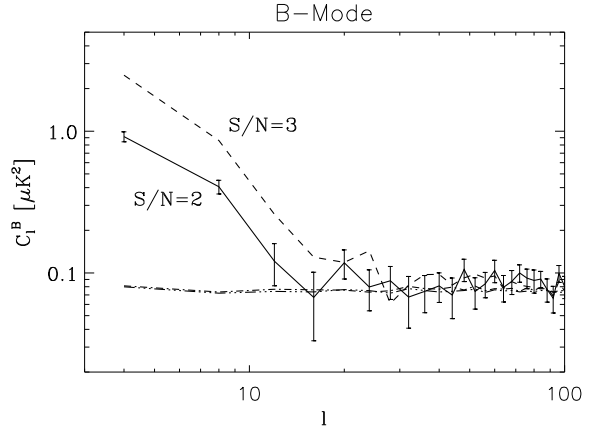}
\caption{$B$-mode spectra 
         of the WMAP 22.8-GHz polarization map in the 
         two regions with $S/N < 2$ (solid) and $S/N < 3$ (dashed). 
         To avoid confusion,
         error bars are reported for the $S/N < 2$ case only. 
         The mean noise spectra of the Montecarlo 
         simulations are also reported (dot--dashed and triple-dot-dashed 
         for $S/N < 2$ and $S/N < 3$, respectively).
\label{spec23Fig}
}
\end{figure}

Figure~\ref{spec23Fig} shows $B$--mode spectra in the two selected areas. 
These are steep on large angular scales, with the
$S/N < 2$ case having less power than $S/N < 3$, as expected according to
selection criteria. A flattening instead occurs on
small angular scales. This behaviour indicates the signal prevails
on large scales 
($\ell < 10$ and $\ell < 15$ for $S/N = 2$ and 3, respectively), 
while the noise dominates on smallest ones.
To check if the flat component is compatible with noise, 
we performed Montecarlo simulations. We have generated
100 noise map realizations assuming the sensitivity map provided 
in the WMAP data package. Then we compute their spectra using only the pixels
of the two selected regions. 
For both the two cases, we find that the average spectrum is flat 
with normalization $C_\ell^{\rm n} = 7.5 \times 10^{-2}$~$\mu$K$^2$ and
$7.3 \times 10^{-2}$~$\mu$K$^2$ for $S/N<2$ and $S/N<3$, respectively,
which are compatible with the level of our spectra at large $\ell$ 
(Figure~\ref{spec23Fig}).
In addition, the spectra of the two $S/N$ cases converge to the same
value in the flat range, which again suggests the spectra are noise dominated
there. We use the mean noise spectra attained with the Montecarlo simulations 
to account for the noise bias in all the next evaluations.

We tried to fit a power law 
\begin{equation}
  C_\ell^B = A^B_{10} \left({\ell\over 10}\right)^{\beta^B}
\end{equation}
to the spectra. Only the case $S/N < 3$ allows meaningful results (Table~\ref{fitTab}),
because only few points (three) can be effectively used for $S/N < 2$,
after the noise bias is subtracted for. 
To provide an estimate of the mean emission 
also for the $S/N < 2$ case, we compute the 
mean value of the quantity $\ell(\ell+1)/(2\pi) C^B_\ell$ over 
the usable $\ell$-range, which is $\ell \le 12$. The result
is reported in Table~\ref{fitTab}, along with the same estimate for the
$S/N<3$ area. It is worth noting that there is a factor $\sim 3$ in spectrum 
($\sim \sqrt{3}$ in signal) between the mean emissions of the two analysed regions.
Moreover, 
the emission we found in the $S/N < 2$ region is a factor $\sim 10$ 
weaker than that measured in the $\sim 75$\% 
sky fraction in the same $\ell$-range \citep{page06}.

\begin{table}
 \centering
  \caption{Mean value of $\left<\ell(\ell+1)/(2\pi)C^B_\ell \right>$
           -- see text. Fit parameters of $B$-mode spectra for $S/N < 3$ are
           also reported. The $\ell$-range used for the 
           estimates is indicated.}
  \begin{tabular}{@{}ccccc@{}}
 \hline
  $S/N$      &   $\ell$-range   &   $A^B_{10}$ [$\mu$K$^2$]    &   $\beta^B$  &   
      $\left<{\ell(\ell+1)\over 2\pi}\,C^B_\ell \right>$ [$\mu$K$^2$]  \\
 \hline
   2        &  [4, 12]         &                  &                  &  $2.8\pm0.3$\\
   3        &  [4, 24]         &   $0.30\pm0.06$  &    $-2.4\pm0.3$  &  $7.62\pm0.14$\\
  \hline
  \end{tabular}
 \label{fitTab}
\end{table}

\section{Implication for CMBP $B$-mode and discussion}\label{CMBSec}

The contamination of the CMB $B$-mode signal can be estimated by extrapolating 
the spectra of Sect.~\ref{specSec} to 70-GHz, a frequency in the range where 
the combined contribution of synchrotron and dust 
is believed to be minimum \citep{page06,carretti06}.
It is worth noting that typical values of rotation measure at high 
Galactic latitudes ($RM \sim 10$--20~rad~m$^{-2}$) generate polarization
angle rotations of 6--12~arcmin at 22.8-GHz. 
Thus, Faraday rotation effects can be considered negligible,
allowing safe frequency extrapolations.

Our spectra are noise dominated
at the multipole $\ell$ the CMB $B$-mode peaks at ($\ell\sim 90$) and
a direct extrapolation of them would provide just upper limits. 
Then, we adopt another approach: we use the spectra measured
in the $\ell$--range where they are signal--dominated and perform angular
extrapolations to the scale of interest.
The mean emission is that provided by the last column of Table~\ref{fitTab}. 
As for the extrapolation, the analysis of the 1.4-GHz DRAO survey data 
shows that the angular behaviour of
 Galactic synchrotron spectra is well represented by a power law 
\begin{equation}
 C_\ell \propto \ell^\beta
 \label{plEq}
\end{equation}
with slope $\beta$ varying in the range [-2.5, -3.0] for $\ell = 2$--300 
\citep{laporta06,burigana06}. 
This is consistent with the value of $\beta=-2.6$ quoted by the WMAP team for the 
whole 75\% of the sky they consider \citep{page06}.
We use as slope both the 
two edge values of the $\beta$ range (-2.5 and -3.0)
and start extrapolations from the middle
of the $\ell$-range used to evaluate the mean spectrum, namely $\ell_0 = 8$,~14
for $S/N < 2$,~$<3$, respectively. 

Frequency extrapolations are performed assuming the brightness Temperature of the
Galactic synchrotron follows a power law $T_b^{\rm synch} \propto \nu^\alpha$, with $\alpha = -3.1$ \citep{bernardi04}.

\begin{figure}
\centering
  \includegraphics[angle=0, width=1.0\hsize]{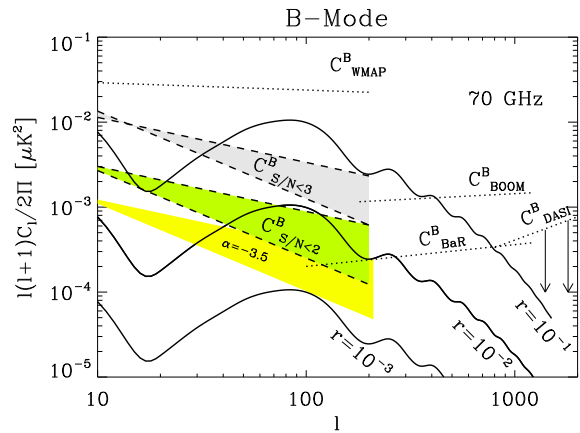}
\caption{$B$--mode power spectra of the Galactic synchrotron emission estimated at 
         70-GHz in the regions with $S/N < 2$ (green shaded area) and $S/N < 3$
         (gray shaded area). 
         The two angular extrapolations with $\beta = -2.5$ and $-3.0$ are
         reported (upper and lower dashed line of shaded areas). A slope
         $\alpha=-3.1$ is assumed for the frequency extrapolation. The case of a
         slope $\alpha=-3.5$ is also reported for the $S/N < 2$ region 
         (yellow shaded).
         For comparison, the plot shows CMB spectra for three
         different values of $r=T/S$, the synchrotron emission detected 
         in the two low emission areas in the target fields of 
         the BOOMERanG
         ($C^B_{\rm BOOM}$, \citealt{carretti05b}) 
         and BaR-SPOrt experiments ($C^B_{\rm BaR}$, \citealt{carretti06}), 
         and the upper limit found in the DASI fields ($C^B_{\rm DASI}$, 
         \citealt{bernardi06}).
         Finally, the general contamination at high Galactic latitude,
         as estimated by the WMAP team using 74.3\% of the sky is
         also reported: the spectrum at 22.8-GHz is considered 
         \citep{page06} and extrapolated to 70-GHz.
\label{spec70Fig}
}
\end{figure}

Results are reported in Figure~\ref{spec70Fig}. 
For the $S/N < 2$ case, which concerns $\sim 16$\% of the sky, 
the contamination range covered by 
the two angular extrapolations at $\ell \sim 90$ competes with the CMBP $B$-mode 
for models with $T/S = [3 \times 10^{-3}$,~$1 \times 10^{-2}]$.
The comparison with previous measurements shows these values are significantly 
lower than the mean contamination estimated by
the WMAP team for most of high Galactic latitudes. 
On the other hand, our results are similar 
to the contamination estimated in the small sky patches observed
in low emission regions (\citealt{carretti06} and references therein)
giving a valuable 
consistency of results in regions selected with similar criteria 
(areas in lowest emission regions).
Although with the uncertainties of an angular scale extrapolation, 
our results extend to the whole lowest emission regions the promising indications
obtained in those few small sample areas covering just few square degrees, 
and strongly support the possibilities to access small $T/S$ values 
in the lowest emission part of the sky.

The use of frequency slope $\alpha = -3.1$ 
can be considered as a conservative approach. 
In fact, \citet{hinshaw06} fit the behaviour of the synchrotron total intensity, 
finding out a steepening of the slope at the highest frequencies of the WMAP range.
In particular, they fit $\alpha=-3.5$ in the range 23--60~GHz. 
With such a slope the synchrotron contamination at 
70~GHz further reduces by a factor 2.5 with respect to $\alpha=-3.1$ and
even $T/S \sim [1 \times 10^{-3}$,~$3 \times 10^{-3}]$ becomes accessible
(yellow shaded area of Figure~\ref{spec70Fig}).

The situation is slightly worse for $S/N < 3$, as expected, since
higher emission pixels are included in the analysis. However, 
$T/S = [2\times10^{-2}$,~$4\times10^{-2}]$ looks still accessible even in this larger
region, assuming $\alpha=-3.1$. These are values better than that
for all the high Galactic latitudes (WMAP results on $\sim 75\%$ of the sky), 
or in the areas
analysed by \citet{laporta06}.

For the lowest emission areas, the minimum combined synchrotron-dust 
contamination has been found to be somewhere between 70 and 90-GHz
\citep{carretti06}. It is thus worthy estimating the synchrotron contamination
of our best case ($S/N < 2$) even at 90-GHz. A further factor 
$\sim 4$ is gained with respect to 70-GHz (in thermodynamic temperature), 
and the Galactic signal competes with models 
with $T/S \sim [1 \times 10^{-3}$,~$3 \times 10^{-3}]$.


Despite the significant general contamination of high Galactic 
latitudes,  the results obtained by our analysis depict a situation with better and
interesting conditions in the lowest emission regions. The possibility to reach
$T/S$ values as low as $10^{-3}$--$10^{-2}$ looks now extended from few small
good sky patches to an area representing $\sim 16$\% of the sky, that
could be the right place for deep CMBP observations
looking at the $B$-mode.

To limit observations in small regions, however, imposes
intrinsic constraints on the minimum detectable $T/S$, 
mainly because of the leakage from $E$- into the weaker $B$-mode.
In fact, \citet{amarie05} find that an all-sky survey would allow a detection 
of $T/S$ with a theoretical sensitivity 
limit of $\Delta(T/S) = 1.5 \times 10^{-5}$, that becomes
$\Delta(T/S) = 3.2 \times 10^{-5}$ when 70\%
of the sky is available, and $\Delta(T/S) = 10^{-3}$ for 15\% (3-$\sigma$ C.L.). 
It is worth noting that the class of inflationary models with minimal fine-tuning 
have $T/S$ values ranging within $10^{-3}$ and $10^{-1}$ \citep{boyle06}, 
for which a 15\% sky portion would be large enough for the first detection of the tensor
CMBP component.

In spite of these limitation, the minimum $T/S$ value detectable in 15\% of the
sky almost matches the limit imposed by the foregrounds we have found 
in lowest emission regions.
Although the uncertainties because of the angular extrapolations we need to apply,
the search for the $B$-mode in a sky portion of such a size could thus be a good
trade-off between intrinsic and foregrounds limits.

In addition, the weakness of the $B$-mode signal makes already a challenge
to detect the signal for $T/S=0.1$ with the present technology 
(e.g. \citealt{cortiglioni06}). 
It is likely that significant technological improvements will be necessary
before cosmologists can face an all-sky mapping mission with a sensitivity 
able to match the intrinsic limit of $\Delta(T/S)\sim 10^{-5}$.
Therefore, an experiment aiming at detecting the $B$-mode in a smaller region 
(10-15\% of the sky) can be a valuable intermediate step that would allow us to
probe inflation models down to $T/S = 10^{-3}$.

\section*{Acknowledgments}

We thank an anonymous referee for valuable comments, which helped improve the paper.
Some of the results in this paper have been derived using the 
HEALPix package (http://healpix.jpl.nasa.gov).
We acknowledge the use of the CMBFAST package, WMAP data, 
and the Legacy Archive for Microwave Background Data Analysis (LAMBDA). 
Support for LAMBDA is provided by the NASA Office of Space Science.

\bsp

\label{lastpage}


\begin{thebibliography}{99}
\bibitem[\protect\citeauthoryear{Amarie et al.}{2005}]{amarie05} 
        Amarie M., Hirata C., \& Seljak U., 2005, Physical Review D, 72, id. 123006
\bibitem[\protect\citeauthoryear{Bernardi et al.}{2004}]{bernardi04} 
        Bernardi G., Carretti E., Fabbri R., Sbarra C., Poppi S., Cortiglioni S., 
        Jonas J.L., 2004, MNRAS, 351, 436
\bibitem[\protect\citeauthoryear{Bernardi et al.}{2006}]{bernardi06}
        Bernardi G., Carretti E., Sault R.J., Cortiglioni S.,
        Poppi S., 2006, MNRAS, 370, 2064
\bibitem[\protect\citeauthoryear{Boyle et al.}{2006}]{boyle06}
        Boyle L.A., Steinhardt P.J., \& Turok N., 2006, Phys. Rev. Lett. 96, 111301 
\bibitem[\protect\citeauthoryear{Burigana et al.}{2006}]{burigana06}
        Burigana C., La Porta L.,  Reich P., Reich W., 2006, Astron. Nachr., 327, 491
\bibitem[\protect\citeauthoryear{Carretti et al.}{2005a}]{carretti05a}
        Carretti E., Bernardi G., Sault R.J., Cortiglioni S., 
        \& Poppi S., 2005a, MNRAS, 358, 1 
\bibitem[\protect\citeauthoryear{Carretti et al.}{2005b}]{carretti05b}
        Carretti E., McConnell D., McClure-Griffiths N.M., Bernardi G., Cortiglioni S., 
        \& Poppi S., 2005b, MNRAS, 360, L10 
\bibitem[\protect\citeauthoryear{Carretti et al.}{2006}]{carretti06}
        Carretti E., Poppi S., Reich W., Reich P., F\"urst E., 
        Bernardi G., Cortiglioni S., Sbarra C., 2006, MNRAS, 367, 132
\bibitem[\protect\citeauthoryear{Cortiglioni \& Carretti}{2006}]{cortiglioni06}
        Cortiglioni S., \& Carretti E., 2006, 
        in Fabbri~R. ed., Cosmic Polarization, 
        in press, astro-ph/0604169
\bibitem[\protect\citeauthoryear{G\'orski et al.}{2005}]{gorski05}
        G\'orski K.M., et al., 2005, ApJ, 622, 759
\bibitem[\protect\citeauthoryear{Hinshaw et al.}{2006}]{hinshaw06} 
        Hinshaw G., et al., 2006, ApJ, submitted, astro-ph/0603451
\bibitem[\protect\citeauthoryear{Kamionkowski \& Kosowsky}{1998}]{ka98} 
        Kamionkowski~M., Kosowsky~A., 1998, PRD, 57, 685
\bibitem[\protect\citeauthoryear{Kinney et al.}{2006}]{kinney06}
        Kinney W.H., Kolb E.W., Melchiorri A., \& Riotto A., 
        2006, astro-ph/0605338
\bibitem[\protect\citeauthoryear{La Porta et al.}{2006}]{laporta06}
        La Porta L.,  Burigana C., Reich W., Reich P., 2006, 
        A\&A Letters, in press, astro-ph/0607300
\bibitem[\protect\citeauthoryear{Martin \& Ringeval}{2006}]{martin06} 
        Martin J., Ringeval C., 2006, JCAP, in press, astro-ph/0605367
\bibitem[\protect\citeauthoryear{Page et al.}{2006}]{page06} 
        Page L., et al., 2006, ApJ, submitted, astro-ph/0603450
\bibitem[\protect\citeauthoryear{Sbarra et al.}{2003}]{sbarra03} 
        Sbarra C., Carretti E., Cortiglioni S., Zannoni M., Fabbri R., 
        Macculi C., Tucci M., 2003, A\&A, 401, 1215
\bibitem[\protect\citeauthoryear{Seljak et al.}{2006}]{seljak06} 
        Seljak U., Slosar A., McDonald P., 2006, astro-ph/0604335
\bibitem[\protect\citeauthoryear{Uyan{\i}ker \& Landecker}{2002}]{uyaniker02}
        Uyaniker~B., Landecker~T.L., 2002, ApJ, 575, 225
\bibitem[\protect\citeauthoryear{Wolleben et al.}{2006}]{wolleben06} 
        Wolleben M., Landecker T.L., Reich W., \& Wielebinski R., 2006, A\&A, 448, 411
\bibitem[\protect\citeauthoryear{Zaldarriaga}{1998}]{zaldarriaga98}
        Zaldarriaga M., 1998, Ph.D. Thesis, M.I.T., astro-ph/9806122
\end{thebibliography}
\end{document}